\documentstyle[sprocl,psfig]{article}

\def\be{\begin{equation}}
\def\ee{\end{equation}}
\def\bea{\begin{eqnarray}}
\def\eea{\end{eqnarray}}
\newcommand{\ltsim}{\ {\raise-0.5ex\hbox{$\buildrel<\over\sim$}}\ }

\newcommand{\gtsim}{\ {\raise-0.5ex\hbox{$\buildrel>\over\sim$}}\ }

\begin{document}


\title{Properties of the Outer Halos of Galaxies from the Study of
Globular Clusters}

\author{Stephen E. Zepf}

\address{Dept.\ of Physics and Astronomy, Michigan State University \\
and Department of Astronomy, Yale University, zepf@pa.msu.edu}


\maketitle\abstracts{ This paper reviews some recent work on the 
properties of the outer halos of galaxies. I particularly focus on
recent and upcoming advances made with the study of globular clusters.
Globular clusters can be observed out to $\sim 100$ kpc from the
centers of galaxies, allowing the study of galactic halos well beyond
the regions probed by many other techniques such as observations
of the integrated light of galaxies. In the few well-studied
cases to date, the study of globular cluster systems has provided
dynamical evidence for dark matter halos around elliptical galaxies, 
and demonstrated kinematic differences between different globular 
cluster populations that shed light on the formation history of 
their host galaxies.}

\section{Why Outer Halos?}

	One of the primary motivations for studying the outer halos
of galaxies is to study the nature of dark matter. 
The reason the outer halos of galaxies
are important for addressing the nature of dark matter is that most galaxies
are only dark matter dominated at large distances from their centers.
In the inner regions of galaxies ($r \ltsim R_{e}$), the observed
stars and gas typically make a significant contribution to the mass budget.
Exactly what that contribution is for different galaxies remains
a subject of active research and vigorous debate. This vigorous
debate is symptomatic of the problem of determining the dark matter
content of the inner parts of galaxies accurately. Specifically,
because the observed baryons make up a non-negligible fraction of 
the mass in the inner regions of most galaxies, an 
accurate determination of the dark matter properties in these regions 
requires a very accurate accounting of the mass contribution 
of these baryons. Since it is a severe challenge to achieve such an accurate 
accounting of the baryonic mass in the inner regions of galaxies,
the dark matter properties of the inner regions of galaxies remain 
somewhat uncertain. Moreover, in regions in which baryons are a significant
fraction of the mass, the dissipation, star formation, and feedback
that can happen in the baryons may affect the distribution of the
dark matter. Although it is possible to study special classes of objects 
(e.g.\ low surface brightness galaxies) for which the observed baryons 
appear to be negligible in the mass budget at all radii,
this approach still leaves open the question of the dark matter
properties of galaxies in general.
An obvious path to take is to study the outer halos of galaxies 
in which the baryonic contribution is much smaller, and one can
obtain a ``clearer'' view of the dark matter around galaxies.

	A second motivation for studying the outer halos of galaxies
is that the outskirts of galaxies have long dynamical times, and
therefore these regions may retain more of a memory of their initial
conditions than central regions for which the crossing times are
much smaller than the Hubble time. Specifically referring to the 
shapes of galaxies and their halos, it is possible that the outer 
regions might more closely reflect the initial conditions, while 
various dissipative processes may re-arrange the matter in the central 
regions.

	A third reason for studying the outer halos of galaxies
is that they are still uncharted territory. Why this is the case, and
some possibilities for making advances in this area are the subject
of the following sections.

\section{Why Globular Clusters?}

	The most significant challenge to studying galaxies at
large distances from their centers is that there is very little
light in the regions to observe. For galaxies rich in neutral
hydrogen, HI disks can be traced to large radii, and these provide
valuable constraints on the dark matter profile at large radii.
These constraints on the dark matter profiles provided by extended
HI rotation curves are some of the strongest available. However,
even in these cases, one would ideally like to constrain the
shape of the dark matter halo in a galaxy and not just its profile.
This has proven to be a very challenging task (see review by Sackett 1999). 

	Early-type galaxies do not have an easily observed tracer such 
as HI, so one needs to look for other ways to constrain the mass
profile of these galaxies. One approach is to determine the
velocity dispersion of the integrated light. However,
as shown in Figure 1, the integrated light of these galaxies drops
rather rapidly with increasing radius, and falls well below the
sky brightness at very modest radii. As a result, even the most 
dedicated attempts with large
telescopes have not been able to constrain the velocity dispersion
of the stars beyond about $2 R_e$, and radial limits of about half
of that are more typical. Data adequate to characterize the higher
order moments useful for constraining the orbital anisotropy are at
least as limited in radial extent. Moreover, one would also like to study
the stellar populations of galaxies at large radii, and the rapid
decline of the surface brightness of the integrated light of all
galaxies makes this a daunting challenge.

\begin{figure}[t]
\hspace{0.5in}
\psfig{figure=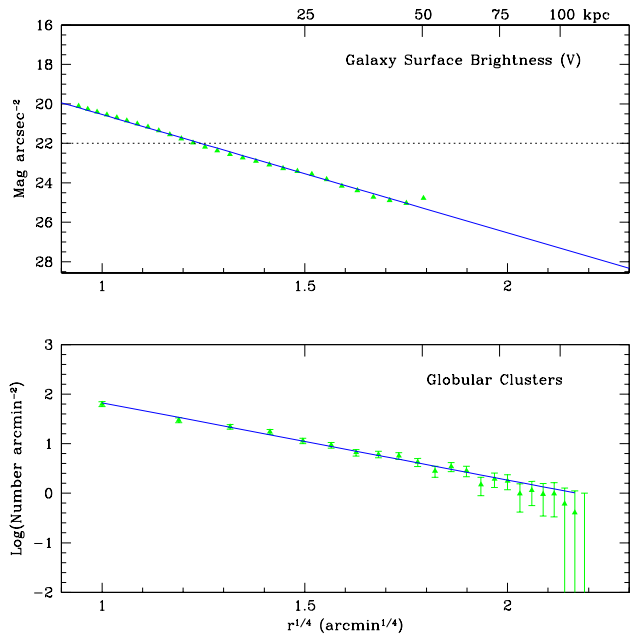,width=9cm}
\vskip-0.08in
\caption{A plot of the surface density against radius of the starlight
(upper panel) and globular clusters (lower panel) for the Virgo
elliptical NGC~4472 (data from Rhode \& Zepf 2001). The light dashed line
represents the surface brightness of a dark sky. This plot demonstrates
the great difficulty extending studies of the integrated light to
large radii. The greater spatial extent of the globular cluster system
relative to the integrated light is also clearly seen.
}
\end{figure}

	Globular clusters provide a valuable tracer of both the
kinematics and the stellar populations of galaxies out to large radii. 
Firstly, as individual dense collections of stars, there is no problem 
observing globulars that are present at large radii. Secondly, as 
shown in Figure 1,
globular clusters have extended spatial distributions, so they can be 
found in useful numbers at large radii, particularly around luminous 
ellipticals which typically have populous globular cluster systems. 
Thirdly, globular cluster systems provide information both about the 
dynamics of the outer halos of galaxies through their radial velocities,
and about the chemical enrichment and possibly even age through their
photometric and spectroscopic properties. 

	Planetary nebulae are also valuable probes of the outer
halos of galaxies. Like globular clusters, each planetary nebulae 
can be observed equally well no matter how far away it is from the center 
of its host galaxy. They also have the advantage that much of their light
is emitted in a single line, so obtaining an accurate radial velocity for
a planetary nebula can be straightforward. However, the surface density of
planetary nebulae does not have the large spatial extent of the globular
cluster systems, and they have not yet provided as much information
about the formation history of galaxies at large radii as globulars.
A final approach, reviewed at this meeting by David Buote, is to
use the properties of the hot gas emitting in X-rays found around luminous
early-type galaxies. With sufficiently accurate X-ray imaging spectroscopy,
it is feasible to determine the dark matter distribution and metal
abundances in hot gas, as well as testing for asymmetries that might
indicate objects for which the assumption of hydrostatic equilibrium
in the gas at large radii is questionable. An obvious goal is to combine
as many approaches as possible, as they each have different sets of
assumptions and possible systematic errors, which might be revealed
through careful intercomparison.

\section{Two Dimensional Shapes and Inferences about Three-Dimensional
Distributions}

	The two dimensional distribution of light in galaxies
has been fairly well characterized within about $1 R_e$.
One of the uses of these data is to try to constrain the three-dimensional 
shapes of galaxies. An unconstrained inversion of two-dimensional data 
to the intrinsic three-dimensional shape is problematic (Rybicki 1987), 
but either through constrained inversion techniques (e.g.\ Ryden 1992, 
Lambas, Maddox, \& Loveday 1992)
or through the addition of kinematic data (e.g.\ Franx, Illingworth, \&
de Zeeuw 1991, Bak \& Statler 2000), some progress can be made.
Overall, the evidence suggests that at least a small amount of triaxiality 
is common, with most galaxies being nearly oblate and a modest fraction 
nearly prolate.

It would be of clear interest
to extend these studies to the outer regions of galaxies
which might be less influenced by evolution and more closely
reflect the conditions when they formed. 
CCD Mosaics covering larger areas are beginning to make
this feasible, although the large surveys of galaxies used
in the statistical studies given above are a long way away.
Both the integrated light and the globular clusters can be studied this way,
with the integrated light offering much more signal, and the
clusters potentially reaching to larger radii, but being limited
by defining the two-dimensional shape with a modest number of points.

As an example of how this work might develop in the
future, we present in Figure 2 our new results for the position
angle and ellipticity of the integrated light and globular clusters
around NGC~4472. The position angle of the globular cluster system
is consistent with that of the integrated light over the same radial
range. The latter qualification matters, since our data confirm
that there is a position angle twist in this galaxy, which can be 
taken as evidence for some level of triaxiality.
The ellipticity of the globular cluster system is marginally smaller
(rounder) than that of the galaxy light, but this requires
confirmation by additional data.
\begin{figure}[h]
\hspace{0.56in}
\psfig{figure=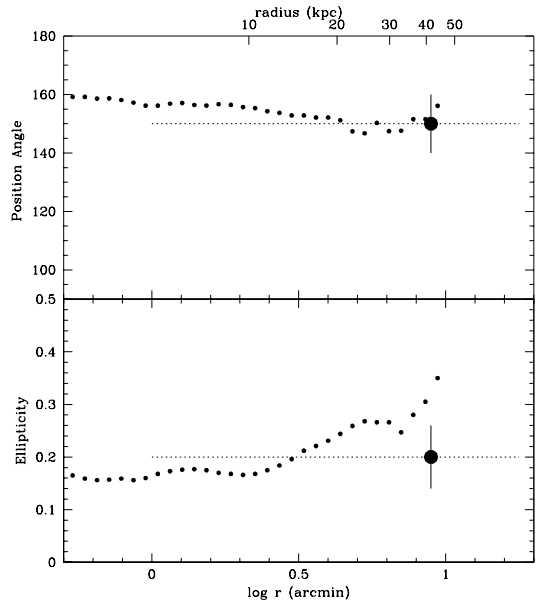,width=8.2cm}
\vskip-0.09in
\caption{The position and ellipticity as a function of radius for
the starlight (plotted as small dots) and the globular clusters
(plotted as a large dot) in NGC~4472. The globular cluster point 
is a radial average over the range indicated by the light dashed line,
with the radial location of the dot as the approximate median radius
of the globular cluster sample. The uncertainty in the position
angle and ellipticity of the NGC~4472 globular clusters system is 
given by the vertical line about the globular cluster point. This
analysis is based on the Mosaic images presented in Rhode \& Zepf (2001).
This plot shows that the position angle of the globular cluster system
is consistent with that of the integrated light over the same radial
range. The ellipticity shows marginal evidence that the globular
cluster system is slightly rounder than the galaxy, but this requires
confirmation by additional data.
\vspace{-0.1truein}  
}
\end{figure}

\section{Radial Velocities}

\subsection{Kinematics of Individual Populations}

\begin{figure}
\vspace{-0.2in}
\hspace{0.4in}
\psfig{figure=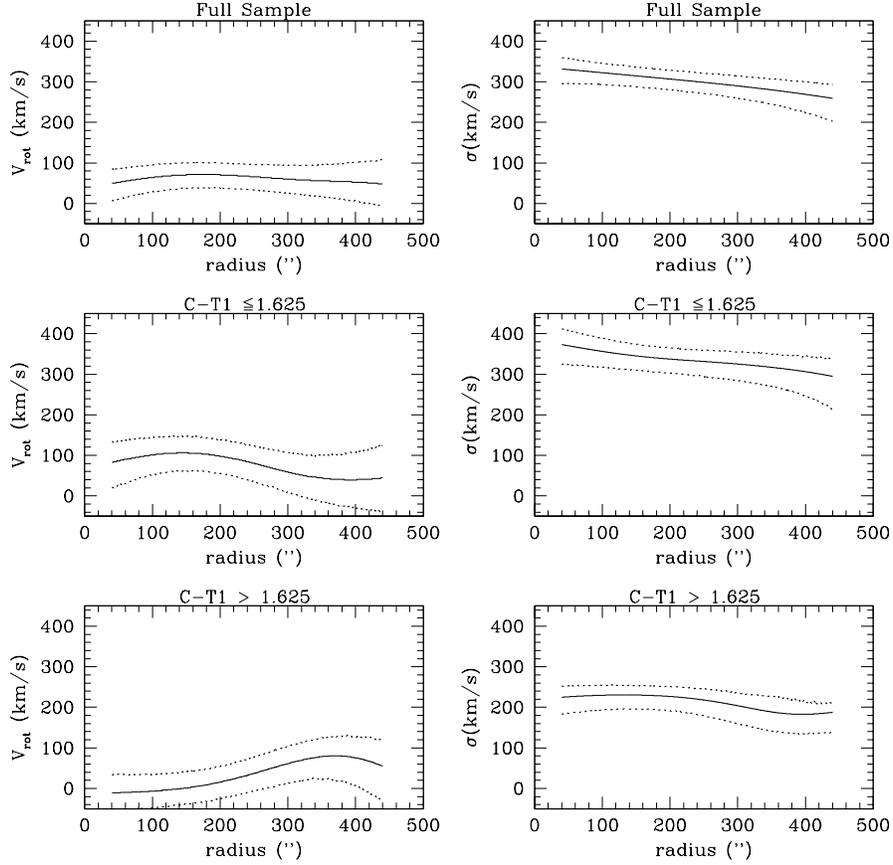,width=12.5cm}
\vskip-0.09in
\caption{Plots of the rotation and velocity dispersion
fields for the globular clusters of NGC~4472 from Zepf et al.\ (2000). 
The top panels are for the full data set, the middle
panels for the metal-poor (blue) clusters and the bottom panels
for the metal-rich (red) clusters. A Gaussian kernel with
$\sigma = 100''$ was used for the radial smoothing for all of
the datasets. The dotted lines show the $1\sigma$ uncertainties,
as determined from bootstrapping. The curves are highly correlated
in the radial direction with the smoothing used.
The plots show modest rotation in the full sample and
the metal-poor cluster population which is essentially constant 
with radius. The red sample has essentially zero rotation
at small radius and a tentative ($1\sigma$) rise to modest
rotation at larger radii. The velocity dispersion
is significantly larger than the rotation at all radii.
\vspace{-0.5truein}  
}
\end{figure}
One of the valuable applications of radial velocities of
substantial numbers of globular clusters around elliptical galaxies
is to compare the kinematics of the metal-rich and metal-poor globular
clusters previously identified in photometric studies. The kinematics
of these systems can shed light on their formation history. 
As an example, results from our recent study of the
NGC~4472 cluster system are shown in Figure 3.
These data confirm our earlier result (Sharples et al.\ 1998) that 
the metal-poor globulars have a larger velocity dispersion than the
metal-rich globulars. Perhaps most importantly, the results shown in
Figure 3 indicate that the metal-rich globular cluster system
has little or no rotation, with an upper limit of $(v/\sigma)_{proj} < 0.34$
($99\%$ confidence level). The absence of rotation in the metal-rich
population in this elliptical strongly distinguishes NGC~4472 
from spirals like those in the Local Group, which have metal-rich
cluster populations with significant rotation. This result argues 
against models in which all metal-rich systems formed more or less 
similarly with the only difference being the mass of the central 
forming ``bulge''.
Instead, the comparison of the significant rotation in the metal-rich
Galactic clusters with the insignificance of rotation in the metal-rich
clusters of NGC~4472 suggests a model in which
elliptical galaxies like NGC~4472 form in major mergers which
create the metal-rich globular cluster population and transfer 
angular momentum outwards, while disk galaxies like the Milky Way 
have only had more minor mergers, which may lead to modest 
amounts of globular cluster formation but which are not as efficient 
at angular momentum transfer.

	It is of interest to compare the results for
NGC~4472 to those of other ellipticals. There are two other galaxies 
for which published data are sufficient to make reliable
statements about the kinematics of their globular cluster systems. One of these
is M87, the central galaxy in the Virgo cluster. Here C\^ot\'e et al.\ 
(2001) used data from Cohen (2000), Cohen \& Ryzhov (1987) and their own
observations to find $(v/\sigma)_{proj} \sim 0.4$, although with large 
error bars because of large uncertainties in the rotation. 
There is evidence that much of the rotation signal comes from
the outer regions (see also Kissler-Patig \& Gebhardt 1998), so 
modest rotation and significant angular momentum transport is 
also suggested for the metal-rich system of this giant elliptical. 
The third system with significant data is the recent merger, 
NGC~5128 (Cen A). The kinematics here appear to be different, 
in that the metal-rich system appears to be rotating significantly, 
while the metal-poor system shows little rotation 
(Hui et al.\ 1995 and references therein).
Possible differences between this system and that of NGC~4472
and M87 are that the NGC~5128 system has not yet come to equilibrium and
transported angular momentum outwards, or that NGC~5128 is a
lower luminosity elliptical which tend to be more rotationally
supported.

\vspace{-0.12in}
\subsection{Mass Profiles}

\begin{figure}[t]
\vspace{-0.72in}
\hspace{0.46in}
\psfig{figure=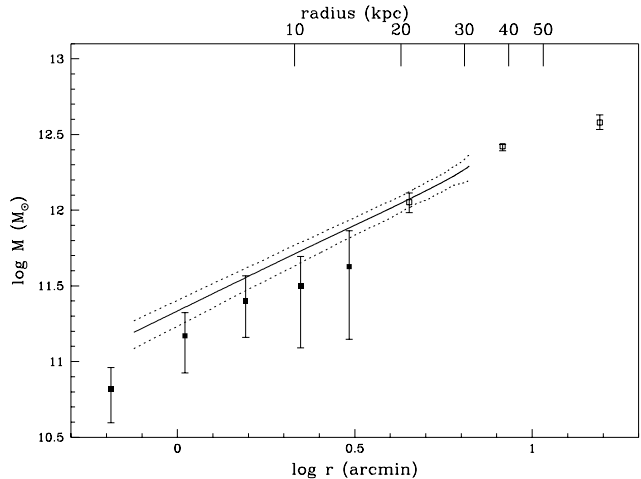,width=10.1cm}
\vskip-0.80in
\caption{A plot of the mass of NGC~4472 as a function of radius
from Zepf et al.\ (2000).
The lines are masses inferred from the radial velocities
of the globular clusters. The central solid line is the best
fit to the 144 radial velocities discussed in this paper.
The dotted lines are the $1\sigma$ lower and upper limits
determined via bootstrapping. 
All of these are based on the assumption of isotropic orbits
for the globular clusters.
The points are masses inferred from ROSAT observations of the 
hot gas around NGC~4472 (Irwin \& Sarazin 1996).
The open squares represent points for which the assumption
of hydrostatic equilibrium on which the X-ray masses are
based may be uncertain because the X-ray isophotes are
irregular at these radii. The overall agreement between
the masses inferred from the two techniques is good,
suggesting that the assumptions underlying each approach
are probably roughly correct. The conclusion that then follows 
is that NGC~4472 has a substantial dark halo, with a mass-to-light 
ratio at several tens of kpc that is at least a factor of five 
greater than in the inner regions of the galaxy.
}
\end{figure}
        Measurements of the radial velocities of globular clusters 
also provide information about the mass distribution
of the host galaxy and the orbits of the clusters. Globular clusters 
are particularly useful probes of the dynamics of the outer halos of 
elliptical galaxies because they can be observed out to much larger 
radii than it is possible to obtain spectroscopy of the integrated light.
A very large number of velocities are required to
independently determine the mass distribution and the orbits
of the tracer particles in a completely non-parametric way
(e.g.\ Merritt \& Tremblay 1994). However, if the mass distribution
inferred from X-ray observations and the assumption of hydrostatic
equilibrium in the hot gas is adopted, the orbits of the
globular clusters can be constrained.
Conversely, if assumptions are made about the cluster
orbits (e.g.\ that they are isotropic), then the mass distribution
can be estimated. In practice, a sensible approach is to
check for consistency of the mass distribution determined
via the X-ray observations of the hot gas with dynamical
measurements (possibly both from globulars and planetary nebulae)
given simplifying assumptions about the orbits, as each technique 
has its own systematic concerns which are
mitigated if the independent approaches agree. 

	In Figure 4, we present the mass profile of NGC~4472,
estimated from the velocity dispersion profile given in Figure 3
and the density profile of the clusters given in Figure 1.
We also show for comparison the mass profile estimate based
on X-ray observations of the hot gas around this galaxy. 
The general agreement in these two mass estimates suggests
both techniques are probably not beset by terrible systematic
errors, and thus provides further evidence for the existence of massive 
dark halos around elliptical galaxies. The agreement between the two
independent mass estimates also suggests that the
assumption of isotropic orbits used to obtain the mass estimate
from the globular cluster velocities is unlikely to be dramatically
off. Similar results are found for M87 (e.g.\ C\^ot\'e et al.\ 2001,
Romanowsky \& Kochanek 2001, Cohen \& Ryzhov 1997). In detail, 
anisotropy is required
at some level for the NGC~4472 globular cluster system because the 
flattening of system can not be supported by the negligible 
rotation observed. This is not necessarily true of the M87
system. One of the potential advantages of the study of individual
tracers (either globular clusters or planetary nebulae) is that
both shape and kinematic information covering the full two-dimensional
projected distribution on the sky are obtained (unlike single slices
from standard long-slit spectroscopy), which may be useful for future
larger studies that attempt to constrain the three-dimensional
shape of galactic halos.

\section*{Acknowledgments}

	The research described here would not have been carried
out successfully without the contributions of many collaborators.
The photometric studies are primarily the thesis work of Katherine Rhode,
and were made feasible by the CCD Mosaic imagers at NOAO. 
The dynamical study of NGC~4472 is based on data obtained at
the CFHT and WHT and came to fruition through the hard work of my
many colleagues on that project, including Mike Beasley, Ray Sharples,
Terry Bridges, and Dave Hanes.
Support for various aspects of the work presented here has been
provided by NASA Long-Term Space Astrophysics grant NAG5-9651, 
by a NASA GSRP Fellowship for K.~Rhode, and by HST NASA grants 
AR-07981 and AR-08755, from the Space Telescope Science Institute, 
operated by AURA, Inc.\ under NASA contract NAS-5-26555.


\section*{References}
\begin{thebibliography}{99}

\bibitem{}J. Bak and T.S. Statler, {\em AJ}, {\bf 120}, 110 (2000).

\bibitem{}J.C. Cohen, {\em AJ}, {\bf 119}, 162 (2000).

\bibitem{}J.C. Cohen, and A. Ryzhov, A., {\em ApJ}, 486, 230 (1997).

\bibitem{}P. C\^ot\'e, {\em et al.}, {\em ApJ}, in press, astro-ph/0106005 
(2001).

\bibitem{}M. Franx, G.D., Illingworth and P.T. de Zeeuw, {\em ApJ}, 
{\bf 383}, 112 (1991).

\bibitem{}M. Kissler-Patig and Gebhardt, K., {\em AJ}, {\bf 116}, 2237, (1998)

\bibitem{}X. Hui, H.C. Ford, K.C. Freeman and M.A. Dopita, {\em ApJ}, 
{\bf 449}, 592 (1995).

\bibitem{}J.A. Irwin and C.L. Sarazin, {\em ApJ}, {\bf 471}, 683 (1996).

\bibitem{}D.G. Lambas, S.J. Maddox and J. Loveday, {\em MNRAS}, {\bf 258},
404 (1992).

\bibitem{}D. Merritt, {\em PASP}, {\bf 111}, 129 (1999).

\bibitem{}D. Merritt and B. Tremblay {\em AJ}, {\bf 108}, 514 (1994).

\bibitem{}K.L. Rhode and S.E. Zepf, {\em AJ}, {\bf 121}, 210 (2001).

\bibitem{}A.J. Romanowsky and C.S. Kochanek, {\em ApJ}, {\bf 552}, 722 (2001).

\bibitem{}G.B. Rybicki in {\em Structure and Dynamics of Elliptical 
Galaxies}, ed.\ P.T. de Zeeuw, 397 (Kluwer, Dordrecht, 1986).

\bibitem{}B.S. Ryden, {\em ApJ}, {\bf 386}, 42 (1992).

\bibitem{}P. Sackett, in {\em Galaxy Dynamics}, ed.\ D. Merrit, J. Sellwood
and M. Valluri, 393, (ASP, San Francisco, 1999).

\bibitem{}R. Sharples, {\it et al.}, {\em AJ}, {\bf 115}, 2337 (1998).

\bibitem{}S.E. Zepf, {\it et al.}, {\em AJ}, {\bf 120}, 2928 (2000).

\end{thebibliography}
\end{document}